\documentclass[copyright,creativecommons,noderivs,noncommercial]{eptcs}
\providecommand{\event}{WLP 2015} 
\usepackage{breakurl}             

\makeatletter

\def\define#1{\@ifnextchar [{\@MYargdef#1}{\@MYargdef#1[0]}}

\def\@MYargdef#1[#2]#3{\@ifdefinable #1{\@MYreargdef#1[#2]{#3}}}

\def\redefine#1{\edef\@tempa{\expandafter\@cdr\string 
  #1\@nil}\@ifundefined{\@tempa}{\@latexerr{\string#1\space undefined}\@ehc
    }{}\@ifnextchar [{\@MYreargdef#1}{\@MYreargdef#1[0]}}

\catcode`\?=11\relax
\def\@MYreargdef#1[#2]#3{\@tempcnta#2\relax\let#1\relax
\edef\@tempa{\def#1}\@tempcntb \@ne
\let\@?@?\relax\@whilenum\@tempcnta>0
\do{\edef\@tempa{\@tempa\@?@?\the\@tempcntb}\advance\@tempcntb \@ne \advance
\@tempcnta \m@ne}\let\@?@?##\@tempa{#3}}
\catcode`\?=12\relax

\makeatother

\renewcommand{\bmod}%
{\mskip-\medmuskip \mkern5mu \mathbin{\idFont mod} \mkern5mu \mskip-\medmuskip}

\define{\mathsym}[1]{\relax\ifmmode#1\else
	\errmessage{Mathematical symbol outside math mode}\fi}
\define{\idFont}{\sf}
\define{\id}[1]{\mathsym{{\idFont #1}}}

\define{\m}[1]{$#1$}
\define{\M}[1]{$$#1$$}

\define{\gt}{\mathsym{\mathchar 12606\relax}} 
\define{\lt}{\mathsym{\mathchar 12604\relax}} 

\define{\paren}[1]{(#1)}
\define{\parenA}[1]{(#1)}
\define{\parenB}[1]{\bigl(#1\bigr)}
\define{\parenC}[1]{\Bigl(#1\Bigr)}
\define{\parenAuto}[1]{\left(#1\right)}
\define{\set}[1]{\lbrace#1\rbrace}
\define{\setA}[1]{\lbrace#1\rbrace}
\define{\setB}[1]{\bigl\lbrace#1\bigr\rbrace}
\define{\setC}[1]{\Bigl\lbrace#1\Bigr\rbrace}
\define{\setCond}[2]{\lbrace#1\mathrel{\vert}#2\rbrace}
\define{\setCondA}[2]{\lbrace#1\mathrel{\vert}#2\rbrace}
\define{\setCondB}[2]{\bigl\lbrace#1\bigm\vert#2\bigr\rbrace}
\define{\setCondC}[2]{\Bigl\lbrace#1\Bigm\vert#2\Bigr\rbrace}

\define{\union}{\cup}
\define{\unionUntil}{\union\cdots\union}
\define{\unionMulti}[2]{\bigcup_{#1}^{#2}}
\define{\intersect}{\cap}
\define{\intersectUntil}{\intersect\cdots\intersect}
\define{\intersectMulti}[2]{\bigcap_{#1}^{#2}}
\define{\timesUntil}{\times\cdots\times}
\define{\setsize}[1]{{\left\vert #1 \right\vert}}
\define{\compl}[1]{\overline{#1}}

\define{\powerset}[1]{2^{#1}}
\define{\inB}{\;\in\;}

\define{\defEq}{:=}
\define{\defEqB}{\;:=\;}
\define{\defIff}{\;\mathsym{:\Longleftrightarrow}\;}
\define{\metaThen}{\mathsym{\;\Longrightarrow\;}}
\define{\metaIf}{\mathsym{\;\Longleftarrow\;}}
\define{\metaIff}{\mathsym{\;\Longleftrightarrow\;}}

\define{\lneg}{\mathsym{\neg}}
\define{\lif}{\mathsym{\leftarrow}}
\define{\lifDup}{\mathsym{\Leftarrow}}
\define{\lthen}{\mathsym{\rightarrow}}
\define{\liff}{\mathsym{\leftrightarrow}}
\define{\lfalse}{\id{false}}
\define{\ltrue}{\id{true}}

\define{\lorUntil}{\lor\cdots\lor}
\define{\landUntil}{\land\cdots\land}
\define{\lorMulti}[2]{\bigvee_{#1}^{#2}}
\define{\landMulti}[2]{\bigwedge_{#1}^{#2}}

\define{\lfor}[4]{#1\:#2\:#3:#4}
\define{\lexistsQ}{\exists}
\define{\lexists}[3]{\lfor{\lexistsQ}{#2}{#1}{#3}}
\define{\lallQ}{\forall}
\define{\lall}[3]{\lfor{\lallQ}{#2}{#1}{#3}}

\define{\answerPred}{\id{answer}}

\define{\natNum}{\mathsym{{\rm l\kern-0.13em N}}}

\define{\realNum}{\mathsym{{\rm I\kern-0.14em R}}}

\define{\struct}[1]{\langle#1\rangle}
\define{\twoCases}[4]{\left\{\begin{array}{l@{\kern10pt}l}
				#1&\mbox{#2}\\#3&\mbox{#4}
				\end{array}\right.}
\define{\until}{, \ldots,}
\define{\seqOf}[1]{#1^*}
\define{\emptySeq}{\epsilon}
\define{\w}{\id{w}}

\define{\code}[1]{{\tt #1}} 

{\begin{center}\begin{tt}\begin{tabular}{@{}l@{}}}%
{\end{tabular}\end{tt}\end{center}}

	{\begin{center}\begin{tt}\begin{tabular}{@{}l@{ }l@{}}}%
	{\end{tabular}\end{tt}\end{center}}

\define{\U}{{\char95}} 
\define{\LT}{{\char60}} 
\define{\GT}{{\char62}} 
\define{\B}{{\char92}} 
\define{\AMP}{{\char38}} 
\define{\D}{{\char36}} 
\define{\SN}{{\char126}} 
\define{\HASH}{{\char35}} 
\define{\Q}{{\char34}} 
\define{\PCT}{{\char37}} 
\define{\LB}{{\char123}} 
\define{\RB}{{\char125}} 
\define{\HAT}{{\char94}} 

\define{\ALPH}{\id{ALP\kern-0.08em H}}
\define{\LOG}{\id{LOG}}
\define{\VARS}{\id{V\kern-0.17em ARS}}
\define{\var}{\mathsym{X}}
\define{\varA}{\id{X}}
\define{\varB}{\id{Y}}
\define{\varC}{\id{Z}}
\define{\const}{\mathsym{c}}
\define{\constA}{\id{a}}
\define{\constB}{\id{b}}
\define{\constC}{\id{c}}
\define{\data}{\mathsym{d}}
\define{\dataSet}{\mathsym{{\cal D}}}
\define{\sig}{\Sigma}
\define{\SORTS}{\id{{\cal S}}}
\define{\sort}{\id{s}}
\define{\PREDS}{\id{{\cal P}}}
\define{\pred}{\id{p}}
\define{\predA}{\id{p}}
\define{\predB}{\id{q}}
\define{\predC}{\id{r}}
\define{\arity}{\id{n}}
\define{\level}{l}
\define{\FUNS}{\id{{\cal F}}}
\define{\fun}{\id{f}}
\define{\args}{\alpha}
\define{\argsB}{\beta}
\define{\argSorts}{\alpha}
\define{\resSort}{\rho}
\define{\interp}{\id{{\cal I}}}
\define{\interpB}{\id{{\cal J}}}
\define{\ass}{\id{{\cal A}}}
\define{\iV}[2]{(#1,#2)}
\define{\eval}[2]{#1\lbrack\kern-0.15em\lbrack#2\rbrack\kern-0.15em\rbrack}
\define{\evalV}[3]{\eval{\iV{#1}{#2}}{#3}}
\define{\varDecl}{\nu}
\define{\TERMS}{\id{T\kern-0.1em E}}
\define{\term}{\id{t}}
\define{\termB}{\id{u}}
\define{\termC}{\id{v}}
\define{\argA}{\id{a}}
\define{\argB}{\id{b}}
\define{\argC}{\id{c}}
\define{\AT}{\id{AT}}
\define{\FO}{\id{FO}}
\define{\fo}{\varphi}
\define{\foB}{\psi}
\define{\fos}{\Phi}
\define{\modify}[3]{#1\langle#2/#3\rangle}
\define{\impl}{\vdash}
\define{\subst}{\theta}
\define{\substB}{\sigma}
\define{\mgu}{\id{mgu}}
\define{\doSubst}[2]{#2#1}
\define{\doSubstB}[2]{(#2)\,#1}
\define{\HU}{\id{{\cal U}}}
\define{\HB}{\id{{\cal B}}}
\define{\HSet}{\id{H}}
\define{\fact}{\id{F}}
\define{\ground}{\id{ground}}
\define{\LAT}{\id{{\cal M}}}
\define{\LATB}{\id{{\cal N}}}

\define{\emptyClause}{{\hbox{%
	\setlength{\unitlength}{0.24ex}%
	\begin{picture}(5,5)(0,0)
	\put(0,0){\line(0,1){5}}
	\put(0,0){\line(1,0){5}}
	\put(0,5){\line(1,0){5}}
	\put(5,0){\line(0,1){5}}
	\end{picture}}}}

\define{\lit}{\id{L}}
\define{\litA}{\id{A}}
\define{\litB}{\id{B}}
\define{\litC}{\id{C}}
\define{\Body}{\mathsym{{\cal B}}}
\define{\F}{\id{F}} 
\define{\ruleFun}[1]{\id{r}_{#1}}

\define{\prog}{\mathsym{{\idFont P}}}
\define{\T}[1]{\mathsym{{\idFont T}}_{#1}}
\define{\TP}{\T{\prog}}
\define{\Tneg}[2]{\mathsym{{\idFont T}}_{#1,#2}}
\define{\lub}{\id{lub}}
\define{\glb}{\id{glb}}
\define{\lfp}{\id{l\kern-0.1em f\kern-0.1em p}}

\define{\I}{\id{{\cal I}}}
\define{\J}{\id{{\cal J}}}

\define{\nf}{\mathop{\id{not\kern0.2em}}}
\define{\nfPred}[1]{\mathop{\id{not{\U}}#1}}

\define{\free}{\id{f}}
\define{\bound}{\id{b}}

\define{\bp}{\id{bp}}
\define{\binding}{\beta}
\define{\bindingSet}{{\cal B}}

\define{\vars}{\mathsym{{\cal X}}}
\define{\varsA}{\mathsym{{\cal X}}}
\define{\varsB}{\mathsym{{\cal Y}}}
\define{\varsC}{\mathsym{{\cal Z}}}

\define{\inputVars}{\id{input}}

\define{\freeVar}{\id{vars}}
\define{\varsOf}{\id{vars}}
\define{\boundPos}{\id{bound}^+}
\define{\boundNeg}{\id{bound}^-}
\define{\unboundPos}{\id{unbound}^+}
\define{\unboundNeg}{\id{unbound}^-}
\define{\err}{\id{err}}

\newcounter{ProgramLine}
\newcounter{FirstLine}

\newenvironment{progTabular}{%
	\addtocounter{ProgramLine}{-1}%
	\setcounter{FirstLine}{1}%
	\ifvmode\vspace{5mm}\fi%
	\begin{list}{\(\bullet\)\hfill}{
		\parskip 3pt plus 1pt
		\labelwidth 0pt
		\labelsep 0pt
		\leftmargin 8mm
		\listparindent \parindent
		\topsep 4mm
		\parsep 3pt plus 1pt
		\itemsep 0pt
		\partopsep 0pt
		\itemindent 0pt
		\rightmargin 0pt
	}
	\item[]
	\begin{tabular}{@{}r@{\hspace{2mm}}l@{}}}%
	{\end{tabular}%
		\end{list}}

\newenvironment{progPart}[1]{%
		\setcounter{ProgramLine}{#1}%
		\begin{progTabular}}%
	{\end{progTabular}}

	{\end{progTabular}%
		\end{tt}}

	{\end{progTabular}%
		\end{tt}}

\define{\tabX}[1]{\ifnum\value{FirstLine}=1\setcounter{FirstLine}{0}\else
	\ifhmode\\[0.5pt]\fi\fi
	\stepcounter{ProgramLine}(\arabic{ProgramLine})&
	\hspace*{#1}}
\define{\tabA}{\tabX{0cm}}
\define{\tabB}{\tabX{8mm}}
\define{\tabC}{\tabX{16mm}}
\define{\tabD}{\tabX{24mm}}
\define{\tabE}{\tabX{32mm}}
\define{\tabF}{\tabX{40mm}}
\define{\tabG}{\tabX{50mm}}
\define{\tabH}{\tabX{60mm}}
\define{\tabBox}[1]{\vrule height0pt depth0pt width0pt\hbox to14mm{#1\ \hfil}}

\define{\IF}{{\bf if }}
\define{\THEN}{{\bf then }}
\define{\ELSE}{{\bf else }}
\define{\FI}{{\bf fi}}
\define{\FOREACH}{{\bf foreach }}
\define{\FOR}{{\bf for }}
\define{\TO}{{\bf to }}
\define{\FROM}{{\bf from }}
\define{\IN}{{\bf in }}
\define{\WITH}{{\bf with }}
\define{\AND}{{\bf and }}
\define{\NOT}{{\bf not }}
\define{\TRUE}{{\bf true}}
\define{\FALSE}{{\bf false}}
\define{\DO}{{\bf do }}
\define{\OD}{{\bf od}}
\define{\WHILE}{{\bf while }}
\define{\BREAK}{{\bf break}}
\define{\PROCEDURE}{{\bf procedure }}
\define{\BEGIN}{{\bf \code{\LB}} }
\define{\END}{{\bf \code{\RB}} }
\define{\RETURN}{{\bf return }}
\define{\LET}{{\bf let }}
\define{\NEW}{{\bf new }}
\define{\COMPUTE}{{\bf compute }}
\define{\APPEND}{{\bf append }}
\define{\PRINT}{{\bf print }}
\define{\BOOL}{{\bf bool }}
\define{\NIL}{{\bf nil }}
\define{\OUTPUT}{{\bf output }}
\define{\INSERT}{{\bf insert }}
\define{\INTO}{{\bf into }}
\define{\CHOOSE}{{\bf choose }}

\define{\COMMENT}[1]{{\rm /\raisebox{-.6ex}{*} #1 \raisebox{-.6ex}{*}/}}

\define{\CPP}{C{\tt ++}}

\define{\activePart}[1]{\id{a}(#1)}
\define{\delayedPart}[1]{\id{d}(#1)}

\define{\lineno}[1]{\hbox to 1.6em{\hfil\m{\lbrack#1\rbrack}}}
\define{\lineref}[1]{\m{\lbrack#1\rbrack}}
\define{\comment}[1]{\mbox{// #1}}

\define{\C}{\mathsym{C}}

\define{\ruleNo}{\rho}

\define{\edbPredA}{\id{e}}
\define{\edbPredB}{\id{r}}
\define{\edbPredC}{\id{s}}

\define{\state}{\mathsym{{\cal S}}}
\define{\Rule}{\mathsym{{R}}}
\define{\db}{\mathsym{{\cal D}}}

\define{\vertices}{\mathsym{{\cal V}}}
\define{\edges}{\mathsym{{\cal E}}}

\define{\called}[2]{(\lineref{#1},\lineref{#2})}

\define{\edge}{\id{edge}}
\define{\pathPred}{\id{path}}
\define{\color}{\id{color}}
\define{\red}{\id{red}}
\define{\mirrorNode}{\id{mirror\_node}}
\define{\nodeA}{\id{a}}
\define{\nodeB}{\id{b}}

\define{\grandparent}{\id{grandparent}}
\define{\parent}{\id{parent}}
\define{\father}{\id{father}}
\define{\mother}{\id{mother}}
\define{\personA}{\id{ann}}
\define{\personB}{\id{betty}}
\define{\personC}{\id{chris}}
\define{\personD}{\id{david}}

\define{\instRel}{\mathsym{\rightarrow_I}}
\define{\redRel}{\mathsym{\rightarrow_R}}
\define{\redRelDB}[1]{\mathsym{\rightarrow_{R(#1)}}}
\define{\anyRel}{\mathsym{\rightarrow}}
\define{\instClosure}{\id{Inst}^+}
\define{\redClosure}[1]{\id{Red}^+_{#1}}
\define{\transRel}{\mathsym{\rightarrow_T}}

\define{\query}{\id{Q}}
\define{\varS}{\id{V}}
\define{\std}{\id{std}}
\define{\sel}{\id{sel}}
\define{\call}{\id{call}}
\define{\csR}{\id{R}}
\define{\csD}{\id{D}}
\define{\predD}{\id{s}}
\define{\relU}{\mapsto_U} 
\define{\relD}{\mapsto_D} 
\define{\relC}{\mapsto_C} 
\define{\relP}{\mapsto_P} 
\define{\relN}{\mapsto_N} 
\define{\relS}{\mapsto_S} 
\define{\relF}{\mapsto_F} 
\define{\relR}{\mapsto_R} 
\define{\relE}{\mapsto_E} 

\title{Bottom-Up Evaluation of Datalog:\\Preliminary Report}

\author{Stefan Brass\quad\quad Heike Stephan
	\institute{Institut f\"ur Informatik,
		Martin-Luther-Universit\"at Halle-Wittenberg,
		Germany}\\
	\email{brass@informatik.uni-halle.de
		\quad\quad stephan@informatik.uni-halle.de}
}

\def\titlerunning{Bottom-Up Evaluation of Datalog: Preliminary Report}
\def\authorrunning{S.~Brass \& H. Stephan}

\begin{document}

\maketitle



\begin{abstract}
Bottom-up evaluation of Datalog has been studied for a long time,
and is standard material in textbooks.
However,
if one actually wants to develop a deductive database system,
it turns out that there are many implementation options.
For instance,
the sequence in which rule instances are applied
is not given.
In this paper,
we study a method
that immediately uses a derived tuple to derive more tuples
(called the Push method).
In this way,
storage space for intermediate results can be reduced. 
The main contribution of our method
is the way in which we minimize the copying of values at runtime,
and 
do much work already at compile-time.
\end{abstract}


\section{Introduction}
\label{secIntro}

The efficient evaluation of queries expressed as logic programs
remains an everlasting problem.
Of course,
big achievements have been made,
but at the same time problem size and complexity grows.
Any further progress can increase the practical applicability
of logic-based, declarative programming.

Our long-term goal is to develop a new deductive database system.
This has many aspects, 
for instance,
language design.
However,
in the current paper, we exclude all special language features,
including negation,
and focus on efficient query evaluation for basic Datalog.

The magic set method is the standard textbook method
for making bottom-up evaluation goal-directed.
Many optimizations have been proposed,
including our own SLDMagic method~\cite{Bra00}
and a method based on Earley deduction~\cite{BS13}.
We assume in the current paper
that such a rewriting of the program has been done,
so we can concentrate on pure bottom-up evaluation.

As we understand it,
bottom-up evaluation is an implementation of the \m{T_P}-operator 
that computes the minimal model of the program.
However,
an implementation is free in the order in which it
applies the rule instances,
while the \m{T_P}-operator first derives all facts
that are derivable with a given set of known facts,
before the derived facts are used (in the next iteration).
Furthermore,
facts do not have to be stored until the end of query evaluation, 
but can be deleted as soon as
all possible derivations using them have been done,
except for the facts that form the answer to the query.
Therefore,
the sequence of rule instance application becomes important.
If one computes predicate by predicate as the standard textbook method,
one of course needs to store the entire extension of the predicates.
However,
if one uses derived tuples immediately,
it might be possible to store only one tuple
of the predicate during the evaluation.
Of course,
for duplicate elimination and termination,
it might still be necessary to store extensions of a few selected predicates.
It is also not given that tuples (facts)
must be represented explicitly
as records or objects in the program.
It suffices if one knows
where the values of single columns (predicate arguments)
can be found.
In this way,
a lot of copying can be saved because
tuples for the rule head
are typically constructed from values
bound in the rule body.
Of course,
one must ensure that the values are not changed
before all usages are finished.

Our plan is to translate Datalog to~\CPP,
and to generate executable code
from the resulting program.
This permits to use existing compilers
for low-level optimizations 
and 
gives an interface for defining built-in predicates.
In~\cite{Bra10CPP},
we already discussed implementation alternatives
for bottom-up evaluation
and did performance comparisons
for a few example programs.
Now we will improve the ``push method'' from that paper
by changing the set of variables used to represent intermediate facts.
This is essential for reducing the amount of copying.
It also enables us to do more precomputation at ``compile time''.

The idea of immediately using derived facts
to derive more facts is not new.
For instance,
variants of semi-naive evaluation have been studied
which work in this way~\cite{Sch93,SU99}.
It also seems to be related to the propagation of updates
to materialized views.
However,
the representation of tuples at runtime
and the code structure is different from~\cite{Sch93}
(and this is essential for the reduction of copying values).
The paper~\cite{SU99} 
translates from a temporal Datalog extension to Prolog,
which makes any efficiency comparison dependend
on implementation details of the used Prolog compiler.
We also believe that the rule application graph
introduced in our paper is a useful concept.
Further
literature about the implementation of deductive database systems
is, for instance,
\cite{Ram93,DMP94,SSW94,SFH96,Liu99Rol,YK00}.
A current commercial deductive DB 
system is
LogicBlox~\cite{GAK12}.
A benchmark collection is
OpenRuleBench~\cite{LFWK09}.

\section{Basic Definitions}

In this paper,
we consider 
basic Datalog,
i.e.~pure Prolog without negation
and without function symbols
(i.e.~terms can only be variables or constants).
We also assume without loss of generality
that all rules have at most two body literals.
The output of our rewriting methods~\cite{Bra00,BS13}
has this property.
(But in any case,
it is no restriction
since one can introduce intermediate predicates.)
Finally,
we require range-restriction (allowedness),
i.e.~all variables in the head of the rule
must also appear in a body literal.
For technical purposes,
we assume that each rule has a unique rule number.


As usual in deductive databases,
we assume that
EDB and IDB predicates are distinguished
(``extensional'' and ``intensional database'').
EDB predicates are defined by facts only,
e.g.~stored in a relational database
or specially formatted files.
Also program input is represented in this way.
IDB predicates are defined by rules.
There is a special IDB-predicate \m{\answerPred}
that only appears in the head of one or more rules.
The task is to compute the extension
of this predicate in the minimal model of the program,
i.e.~all derivable \m{\answerPred}-facts.

We assume that the logic program for the IDB predicates
as well as the query (i.e.~the \m{\answerPred}-rules)
are given at ``compile time'',
whereas the database for the EDB predicates is only known at ``runtime''.
Since the same program can be executed several times
with different database states,
any optimization or precomputation
we can do at compile time will pay off in most cases.
It might even be advantageous in a single execution
because the database is large. 

Since we want to generate \CPP~code,
we assume that 
a data type known
for every argument of an EDB predicate.
The method does not need type information for IDB predicates
(this is implicitly computed).
%
Data structures for storing relations for EDB predicates
can be characterized with binding patterns:
A binding pattern for a predicate~\m{\pred} with \m{n}~arguments
is a string of length~\m{n} over the alphabet~\m{\set{\bound,\free}}.
The letter~\m{\bound} (``bound'') means that a value for the corresponding
argument is known when the predicate relation is accessed
(input),
\m{\free} (``free'') means that a value needs to be looked up 
(output).

As mentioned above,
our rewriting methods~\cite{Bra00,BS13}
produce rules that have at most two body literals.
Furthermore the case of two IDB-literals
is rare --- it is only used in special cases
for translating complex recursions.
Most rules have one body literal with IDB-predicate
and one with EDB-predicate.
Of course,
there are also rules with only one body literal (EDB or IDB).

\section{Accessing Database Relations}
\label{secDatabaseInterface}

The approach we want to follow
is to translate Datalog into \CPP,
which can then be compiled to machine code.
%
Of course,
we need an interface to access relations for the EDB predicates.
These relations can be stored in a standard relational database,
but it is also possible to program this part oneself
(at the moment,
we do not consider concurrent updates
and multi-user access).

We assume that
it is possible to open a cursor (scan, iterator) over the relation,
which permits to loop over all tuples.
We assume that for every EDB predicate~\m{p}
there is a class \code{\m{p}{\U}cursor} with the following methods:
\begin{itemize}
\item
\code{void open()}: Open a scan over the relation,
i.e.~place the cursor before the first tuple.
\item
\code{bool fetch()}: Move the cursor to the next tuple.
This function must also be called to access the first tuple.
It returns \code{true} if there is a first/next tuple,
or \code{false} if the cursor is at the end of the relation.
\item
\code{\m{T} col{\U}\m{i}()}: Get the value of the \m{i}-th column (attribute)
in the current tuple.
Here \m{T} is the type of the \m{i}-th column.
\item
\code{close()}: Close the cursor.
\end{itemize}
For recursive rules,
we will also need
\begin{itemize}
\item
\code{push()}: Save the state of the cursor on a global stack.
\item
\code{pop()}: Restore the state of the cursor.
\end{itemize}
A relation may have special access structures
(e.g.~it might be stored in a B-tree, hash table or array).
Then not only a full scan
(corresponding to binding pattern~\m{\free\free\ldots\free})
is possible,
but also scans only over tuples with 
given values
for certain arguments.
We assume that in such cases
there are additional cursor classes
called \code{\m{p}{\U}cursor{\U}\m{\binding}},
with a binding pattern~\m{\binding}.
These classes have the same methods as the other cursor classes,
only the \code{open}-method has parameters for the bound arguments.
E.g.~if \code{p} is a predicate 
of arity~3
that permits 
particularly fast access to tuples
with a given value of the first argument,
and if this argument has type~\code{int},
the class \code{p{\U}cursor{\U}bff}
would have the method~\code{open(int x)}.

\section{Duplicate Elimination and Termination}
\label{secDuplicateElimination}


The main contribution of this paper is the way
in which copying and materialization of tuples is avoided.
Our method basically pushes newly derived facts
to body literals where they can be used to derive further facts.

However,
in the presence of recursion,
we must be able to notice
whether a derived tuple is new or not.
Therefore, in each recursive cycle,
at least one predicate must be materialized (``tabled'')
to ensure termination.
A simple solution is to create 
hash tables
for the predicates in question.

This solution means that we 
materialize
the extensions of some IDB predicates
(hopefully, only a few)
and copy all data values for the tuples of these predicates.
In some cases,
information about order or acyclicity
might help to avoid this.
Information about keys and data distribution
could be used to make sensible optimization decisions.
Furthermore,
if tuples are produced in a sort order,
the duplicate check can be done very efficiently
and without storing the predicate extension.
All this is subject of our future work.

It is also interesting
that the data values in a derived tuple
are stored at different times in program variables.
For instance,
we might know that when \m{\pred(\varA,\varB)} is generated,
\m{\varA} only seldom changes,
and \m{\varB} changes much more often.
Then a nested relation might be best for tabling the predicate
for the purpose of duplicate detection.

Of course,
breaking each recursive cycle
with a duplicate detection
is only the minimum we have to do to ensure termination.
Also non-recursive rules can generate duplicates,
and in some cases it might be more efficient to detect these duplicates early
in order to avoid duplicate computations
(since the price for duplicate detection is quite high,
in other cases it might be more efficient
to simply do the duplicate work).

\section{Code Generation: Overall Structure}
\label{secCodeGeneration}

The result of the translation
looks basically 
as shown in Figure~\ref{figOverallStructure}.
\begin{figure}
\begin{quote}
\begin{tt}
\begin{tabular}{@{}l@{}}
	\m{\langle}Declaration Section\m{\rangle};\\
	\m{\langle}Initialization Section\m{\rangle};
		// Initializes backtrack{\U}stack\\
	while(!backtrack{\U}stack.is{\U}empty()) {\LB}\\
	~~~	switch(backtrack{\U}stack.pop()) {\LB}\\
	~~~	~~~	case L1:\\
	~~~	~~~	l1:\\
	~~~	~~~	~~~	\m{\langle}Code Piece 1\m{\rangle};\\
	~~~	~~~	~~~	// break or goto at end of Code Piece\\
	~~~	~~~	case L2:\\
	~~~	~~~	l2:\\
	~~~	~~~	~~~	\m{\langle}Code Piece 2\m{\rangle};\\
	~~~	~~~	...\\
	~~~	{\RB}\\
	{\RB}
\end{tabular}
\end{tt}
\end{quote}
\caption{Overall structure of the generated code}
\label{figOverallStructure}
\end{figure}
So there are many small code pieces,
each with a label
that is suitable for a \code{goto}.
Furthermore,
when there are several things to do,
e.g.~a generated fact can be used in more than one rule, 
a backtrack point is set up for the second rule, 
and then a \code{goto} is done for the first.
When an execution path reaches an end,
the \code{switch} is left with \code{break},
and one of the delayed tasks is taken from the stack.
Therefore,
each code piece also has a unique number,
which can be stored on the backtrack stack,
and used in the \code{switch} to reach the code piece.

Optimizations are possible,
e.g.~one can order the code pieces such that some jumps can be eliminated,
because the target 
is immediately following.
Some backtrack points can be avoided by finding a suitable code sequence.

\subsection{Declaration Section}

\label{subsecDeclaration}

Data not known at compile time
always originates from 
the database.
In order to minimize copying,
we (usually) introduce a \CPP~variable
only for Datalog variables 
which
\begin{itemize}
\item
occur in an EDB body literal, 
\item
but do not occur in an IDB body literal of that rule
(because then the value comes from another rule,
where a variable has been created,
if the value is not known at compile time),
\item
and occur in the head of that rule
(because otherwise the value does not really have to be processed
in the program).
\end{itemize}
For instance,
consider the following rule:
\begin{quote}
\m{\pred(\varA,\varB,\constA)\lif
	\predB(\varB)\land\predC(\varA,\varB,\varC,\varC).}
\end{quote}
If \m{\predB} is an IDB predicate
and \m{\predC} an EDB predicate,
we create a \CPP~variable only for~\m{\varA}.
A variable or constant for~\m{\varB} exists already
when the rule is activated.

In seldom cases of recursive rule applications
(see Section~\ref{subsecVariableConflicts} below)
we create \CPP~variables for all variables of the rule.

If the above condition shows
the we must create a \CPP~variable
for variable~\m{\varA} in rule~\m{\ruleNo},
we generate the following code line in the declaration section:
\begin{quote}
\begin{tt}
\begin{tabular}{@{}l@{}}
\m{T} v\m{\ruleNo}{\U}\m{\varA};\\
\end{tabular}
\end{tt}
\end{quote}
We use the prefix with the rule number so that there can be no name conflicts
between variables of different rules.
\code{T} is the \CPP~data type
for the database column in which \m{\varA} occurs.

\subsection{Symbolic Facts}

A symbolic fact
consists of an IDB~predicate~\m{\pred}
and a tuple~\m{(\term_1\until\term_n)}
of \CPP~variables (i.e.~their identifiers) and constants,
where \m{n} is the arity of~\m{\pred}.
So a symbolic fact represents
what is known at compile time
about a fact that will be derived at runtime.
For some arguments,
we might know the exact value 
(a constant),
for other arguments,
we know the \CPP~variable
which will contain the value.

An initial set of symbolic facts
is derived by rules without IDB body literals.
Then our task is to pass each derived symbolic fact
to matching IDB body literals
and to derive a symbolic fact for the rule head.
For each such rule application,
a code piece is generated
which does the remaining computation at runtime.
The computation of symbolic facts is similar to the standard fixpoint iteration
to compute the minimal model
(but it is done at ``compile time'',
when the data for the EDB predicates are not yet known).

``Matching'' between a symbolic fact and a body literal
means that they are unifiable.
In general a full unification must be done (at compile time).
Consider e.g.~the body literal~\m{\pred(\varA,\varA,\constA)}
and the symbolic fact~\m{\pred(\constB,\code{v1{\U}}\varB,\code{v1{\U}}\varB)}.
The rule cannot be applied to the symbolic fact,
so no code is generated for this case.

\subsection{Rule Application Graph}
\label{subsecRuleApplicationGraph}

As explained above,
we assume that all rules have at most two body literals.
A ``Symbolic Rule Application''
is represented by
\begin{itemize}
\item
a rule from the logic program with one IDB body literal,
together with a symbolic fact matching this body literal, or
\item
a rule without IDB body literals, or
\item
a rule with two IDB body literals,
with one of the two selected,
together with a symbolic fact matching this body literal.
(In the rare case of two IDB body literals,
we use temporary tables for facts matching each body literal.
The symbolic fact in this rule application describes the situation
that we just computed a new fact for one of the IDB body literals.
For the other body literal we use the table with previously computed facts.)
\end{itemize}
The result of a symbolic rule application
is a symbolic fact.
Let \m{\pred(\term_1\until\term_n)}
be the head of the rule,
and \m{\ruleNo} be its rule number.
If the rule has an IDB body literal,
let \m{\subst} be a most general unifier with the input symbolic fact.
We require that variable-to-variable bindings are done such that
logic variables are replaced by \CPP~variables.
Then the derived symbolic fact is \m{\pred(\termB_1\until\termB_n)},
where \m{\termB_i} is
\begin{itemize}
\item
\m{\term_i} if \m{\term_i} is a constant.
\item
\m{\doSubst{\subst}{\term_i}}
if \m{\term_i} is a variable which appears in the IDB body literal
(if there is one).
\item
\m{\code{v}\ruleNo\code{\U}\varA}
if \m{\term_i} is a variable~\m{\varA}
which does not appear in the IDB body literal.
\end{itemize}
Now we can do a standard fixpoint computation
to compute all symbolic facts
which are derivable from the program.
This process will come to an end,
because the number of symbolic facts is bounded:
There is only a finite number of \CPP~variables
(at most the number of variables in the given logic program,
where variables with the same name in different rules count as distinct).
Furthermore,
only a finite number of constants occurs in the given logic program
(constants which appear only in the database
are not known at ``compile time''
and not used for computing symbolic facts).

The structure of the computation can be shown in a
``rule application graph''.
It has two types of nodes,
namely symbolic facts (``fact nodes''),
and symbolic rule applications (``rule nodes'').
There is an edge from every symbolic fact
to every symbolic rule application
which uses the symbolic fact.
Furthermore,
there is an edge from every symbolic rule application
to the symbolic fact it generates.

Of course,
it is possible to show only the rule
in nodes for symbolic rule applications
(since the symbolic fact is identified by the incoming edge,
except in the case of two IDB body literals).
However,
then there can be several nodes marked with the same rule:
It is possible that a single rule
is compiled several times for different symbolic facts
matching its IDB body literal.

Note also that not every application of a recursive rule
to a symbolic fact is actually recursive:
Only if the same symbolic fact can be generated by applying this rule
(maybe indirectly via other rules),
we have to be prepared for recursive invocations
of the code piece for the symbolic rule application.
This can be seen from cycles in the graph.

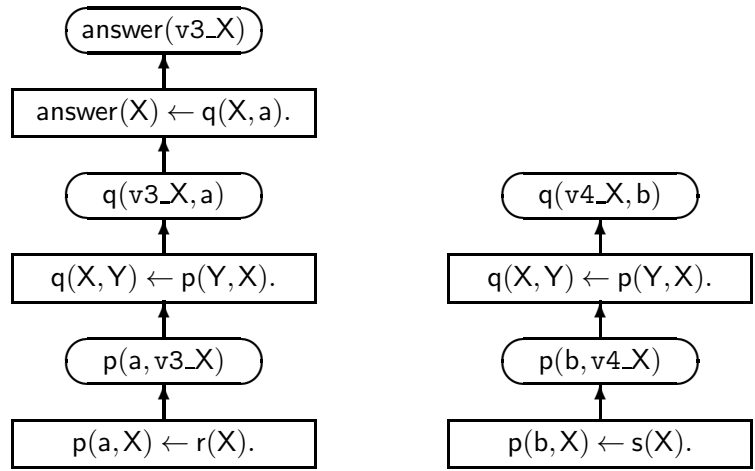
\begin{figure}
\begin{center}
\thicklines
\setlength{\unitlength}{1mm}
\begin{picture}(98,61)(0,0)
\put(  0, 0){\framebox(40,6){\m{\pred(\constA,\varA)\lif\predC(\varA).}}}
\put( 20, 6){\vector(0,1){5}}
\put( 20,14){\makebox(0,0){\m{\pred(\constA,\code{v3{\U}}\varA)}}}
\put( 20,14){\oval(26,6)}
\put( 20,17){\vector(0,1){5}}
\put(  0,22){\framebox(40,6){\m{\predB(\varA,\varB)\lif\pred(\varB,\varA).}}}
\put( 20,28){\vector(0,1){5}}
\put( 20,36){\makebox(0,0){\m{\predB(\code{v3{\U}}\varA,\constA)}}}
\put( 20,36){\oval(26,6)}
\put( 20,39){\vector(0,1){5}}
\put(  0,44){\framebox(40,6){\m{\answerPred(\varA)\lif\predB(\varA,\constA).}}}
\put( 20,50){\vector(0,1){5}}
\put( 20,58){\makebox(0,0){\m{\answerPred(\code{v3{\U}}\varA)}}}
\put( 20,58){\oval(26,6)}
\put( 58, 0){\framebox(40,6){\m{\pred(\constB,\varA)\lif\predD(\varA).}}}
\put( 78, 6){\vector(0,1){5}}
\put( 78,14){\makebox(0,0){\m{\pred(\constB,\code{v4{\U}}\varA)}}}
\put( 78,14){\oval(26,6)}
\put( 78,17){\vector(0,1){5}}
\put( 58,22){\framebox(40,6){\m{\predB(\varA,\varB)\lif\pred(\varB,\varA).}}}
\put( 78,28){\vector(0,1){5}}
\put( 78,36){\makebox(0,0){\m{\predB(\code{v4{\U}}\varA,\constB)}}}
\put( 78,36){\oval(26,6)}
\end{picture}
\end{center}
\caption{Rule Application Graph with Useless Part (to be eliminated).}
\label{figRuleApplicationGraph}
\end{figure}

Finally,
nodes in the graph from which there is no path
to an \m{\answerPred}-node
can be eliminated:
They do not contribute to the computation of the answer.
If the program is the result of a program transformation like magic sets,
this path will not be followed at runtime,
but it is better not to generate code for it.
An example of such a program is
\begin{quote}
	\m{\begin{array}{@{}lcl@{}}
		\answerPred(\varA)&\lif&\predB(\varA,\constA).\\
		\predB(\varA,\varB)&\lif&\pred(\varB,\varA).\\
		\pred(\constA,\varA)&\lif&\predC(\varA).\\
		\pred(\constB,\varA)&\lif&\predD(\varA).\\
	\end{array}}
\end{quote}
The rule application graph is shown in Figure~\ref{figRuleApplicationGraph}.
The right path is useless.
In the code generation below,
we assume that such useless computation paths have been removed,
i.e.~from every node a fact node with predicate~\m{\answerPred}
is reachable.
This in particular means
that every fact node with a predicate different from ``\m{\answerPred}''
has an outgoing edge.


\subsection{Variable Conflicts}

\label{subsecVariableConflicts}

In rare cases of recursive rule applications,
it is possible that a rule is applied to a symbolic fact
which contains already a variable generated for that rule.
An example is
\begin{quote}
\m{\begin{array}{@{}lcl@{}}
\pred(\varA,\varB)&\lif&\predC(\varA,\varB).\\
\pred(\varB,\varC)&\lif&\pred(\varA,\varB)\land\predC(\varB,\varC).
\end{array}}
\end{quote}
The first rule generates the symbolic fact
\m{\pred(\code{v1{\U}}\varA,\code{v1{\U}}\varB)}.
When we insert this into the second rule,
we get
\m{\pred(\code{v1{\U}}\varB,\code{v2{\U}}\varC)}.
Now we have to insert this again into the second rule:
\m{\code{v2{\U}}\varC} contains the input value for~\m{\varB},
but must also be set with a new data value from~\m{\predC}.
In this case,
some copying seems unavoidable.
While there are optimizations possible,
the simplest solution is to create a \CPP~variable
for each logical variable of the rule,
and to copy first the values from the input fact to the right variable
(which might need temporary variables,
e.g.~for swapping the values of two variables).
For recursive rule applications,
the previous variable values are also stored on a stack
(see Section~\ref{subsecProtection} below).

\subsection{Labels for Code Pieces}

We need a \code{goto} label and/or
a \code{case} selector value (a unique number)
for each code piece implementing a symbolic rule application.
We write
\begin{quote}
\code{l{\U}start(\m{\pred(\term_1\until\term_n)},~\m{\ruleNo},~%
	\m{\pred(\termB_1\until\termB_n)})}
\end{quote}
for the \code{goto}-label
of the code piece for application of rule~\m{\ruleNo}
with body literal~\m{\pred(\termB_1\until\termB_n)}
to the symbolic fact~\m{\pred(\term_1\until\term_n)}.
Of course, instead of listing
the body literal \m{\pred(\termB_1\until\termB_n)} explicitly,
one could also use its position number in the rule~\m{\ruleNo}.
In any case,
the implementation will replace this by
\code{l{\U}start{\U}\m{n}}
with some unique number~\m{n}.
The symbolic constant for the \code{case}-value is written as
\code{L{\U}START(\ldots)}
(and also made a legal \CPP~identifier
by using the same unique number).
Sometimes there are continuations
or other code pieces,
therefore the label is marked as ``\code{start}''.


%
%

\subsection{Protection of Variable Values}
\label{subsecProtection}

Of course,
when a code piece
corresponding to 
a symbolic rule application
is executed,
the \CPP~variables in the symbolic fact \m{\pred(\term_1\until\term_n)}
must still have the same value
as when this task was generated.
It is possible that the ID/label of the 
code piece
was pushed on the backtrack stack
and it is executed only later.

However,
for every \CPP~variable,
a new value is assigned only
in code pieces for the single rule
for which the variable was introduced
(to hold a data value for an EDB literal in that rule).

Furthermore,
it is important that the backtrack points are kept on a stack.
So we will return to that rule only after all backtrack points
which use the value (and are thus generated later)
have been processed---unless the rule is recursive.
In this case,
the variable value must be saved
(on another stack suitable for the data type),
and we put the ID of a code piece on the backtrack stack
which restores the variable value.
This is done whenever we enter a recursive rule,
and only for variables set in this rule
(the derived fact might contain also variables passed from elsewhere
and not changed in the rule).

If the backtrack stack shrinks below this point,
all usages of the new variable value are done,
and the old value is restored,
so that older backtrack points find the value
which was current when the backtrack point was created.

\section{Code Pieces}
\label{secCodePieces}

In this section,
we define a number of code pieces
which are translations of different types of rules.
Each code piece corresponds to a symbolic rule application.
For simplicity,
we do not consider variable conflicts (Section~\ref{subsecVariableConflicts})
here.

\subsection{IDB-Facts}

Suppose the program contains an IDB-fact~\m{\pred(\const_1\until\const_n)}.
For each body literal~\m{\pred(\term_1\until\term_n)}
of a rule~\m{\ruleNo}
that unifies with the fact~\m{\pred(\const_1\until\const_n)},
the case selector value
\begin{quote}
\code{L{\U}START(\m{\pred(\const_1\until\const_n)},~\m{\ruleNo},~%
	\m{\pred(\term_1\until\term_n)})}
\end{quote}
is pushed on the backtrack stack
during initialization.

\subsection{One EDB-Body Literal}

\label{subsecOneEDBLit}

Consider the rule
\m{\pred(\term_1\until\term_n)\lif\predC(\termB_1\until\termB_m)}
where \m{\predC} is an EDB predicate.
Let \m{\ruleNo} be the rule number.
Let \m{\pred(\bar\term_1\until\bar\term_n)}
be the symbolic fact generated by the rule
(\m{\bar\term_i\defEq\term_i} if \m{\term_i} is a constant,
and
\m{\bar\term_i\defEq\:}\code{v\m{\ruleNo}{\U}\m{\varA}}
if \m{\term_i} is the variable~\m{\varA}).


Among all possible cursors~\code{cursor{\U}\m{\predC}{\U}\m{\binding}}
for~\m{\predC}
choose one such that for all bound argument positions~\m{i}
(i.e.~\m{\binding_i=\bound}),
\m{\termB_i} is a constant.
This is always possible because every relation supports a full table scan,
i.e.~an access path with all argument positions ``free''.
But obviously,
if there are constants among the~\m{\termB_i},
and there are available indexes,
it is best to choose one with the smallest estimated result size.
In the declaration section,
generate
\begin{quote}
\begin{tt}
\begin{tabular}{@{}l@{}}
cursor{\U}\m{\predC}{\U}\m{\binding} c\m{\ruleNo};\\
\end{tabular}
\end{tt}
\end{quote}
Define symbolic constants \code{L{\U}INIT{\U}\m{\ruleNo}}
and \code{L{\U}CONT{\U}\m{\ruleNo}}
as unique numbers for cases in the switch.
Generate the following code in the initialization section:
\begin{quote}
\begin{tt}
\begin{tabular}{@{}l@{}}
backtrack{\U}stack.push(L{\U}INIT{\U}\m{\ruleNo});
\end{tabular}
\end{tt}
\end{quote}
All following code is generated in the \code{switch}:
\begin{enumerate}
\item
Generate
\begin{quote}
\begin{tt}
\begin{tabular}{@{}l@{}}
case L{\U}INIT{\U}\m{\ruleNo}:
\end{tabular}
\end{tt}
\end{quote}
\item
Let \m{i_1\until i_k} be the bound argument positions in~\m{\binding}.
Generate: 
\begin{quote}
\begin{tt}
\begin{tabular}{@{}l@{}}
~~~ c\m{\ruleNo}.open(\m{\termB_{i_1}\until\termB_{i_k}});
\end{tabular}
\end{tt}
\end{quote}
(Note that although another \code{case} follows,
execution simply continues.).
\item
Generate:
\begin{quote}
\begin{tt}
\begin{tabular}{@{}l@{}}
case L{\U}CONT{\U}\m{\ruleNo}:
\end{tabular}
\end{tt}
\end{quote}
The following loop (item~\ref{EDBLitWhile})
is left with \code{goto}
when the first fact is generated.
But before the jump,
this case label is pushed on the backtrack stack,
so that the loop is continued later.
\item
\label{EDBLitWhile}
Generate
\begin{quote}
\begin{tt}
\begin{tabular}{@{}l@{}}
~~~ while(c\m{\ruleNo}.fetch()) {\LB}
\end{tabular}
\end{tt}
\end{quote}
\item
Let \m{\termB_{i_1}\until\termB_{i_k}}
be the constants among the~\m{\termB_1\until\termB_m}
which correspond to free argument positions in~\m{\binding}.
If \m{k\ge1},
generate
\begin{quote}
\begin{tt}
\begin{tabular}{@{}l@{}}
~~~ ~~~ if(c\m{\ruleNo}.col{\U}\m{i_1}() != \m{\termB_{i_1}} || \m{\cdots}
|| c\m{\ruleNo}.col{\U}\m{i_k}() != \m{\termB_{i_k}})\\
~~~ ~~~ ~~~ continue;
\end{tabular}
\end{tt}
\end{quote}
I.e.~if the current tuple of the EDB-predicate does not have the required
values for the constant arguments,
we immediately start the next iteration of the \code{while}-loop
(i.e.~fetch the next tuple).
\item
For every variable~\m{\varB},
which appears more than once among the~\m{\termB_1\until\termB_m}:
Let \m{\termB_{i_1}\until\termB_{i_k}}
be all equal to~\m{\varB}
(note that \m{k\ge 2}).
Generate a test that the same value appears in these columns:
\begin{quote}
\begin{tt}
\begin{tabular}{@{}l@{}}
~~~ ~~~ if(c\m{\ruleNo}.col{\U}\m{i_1}() \kern-2pt != \kern-2pt
		c\m{\ruleNo}.col{\U}\m{i_2}
	\kern-2pt || \kern-2pt \m{\cdots} \kern-2pt || \kern-2pt
	c\m{\ruleNo}.col{\U}\m{i_{k-1}}() \kern-2pt != \kern-2pt
			c\m{\ruleNo}.col{\U}\m{i_k}())\kern-17pt\\
~~~ ~~~ ~~~ continue;
\end{tabular}
\end{tt}
\end{quote}
\item
For every variable~\m{\varA_i} in the head
let \m{\termB_j} be any occurrence of this variable
among the \m{\termB_1\until\termB_m}.
Because of the range restriction (allowedness) condition on the rules,
\m{\varA_i} must occur in the body.
Generate for each~\m{\varA_i}:
\begin{quote}
\begin{tt}
\begin{tabular}{@{}l@{}}
~~~ ~~~ v\m{\ruleNo}{\U}\m{\varA_i} = c\m{\ruleNo}.col{\U}\m{j}();
\end{tabular}
\end{tt}
\end{quote}
\item
\label{OneEDBLitDuplicateCheck}
In case the predicate~\m{\pred} was selected for a duplicate check,
the following must be done here:
The result tuple
\m{\pred(\bar\term_1\until\bar\term_n)}
with the current values of the \CPP~variables
is entered into a hash table or other data structure.
If the tuple was already present,
one simply does ``\code{continue;}''
to skip it.
\item
Generate:
\begin{quote}
\begin{tt}
\begin{tabular}{@{}l@{}}
~~~ ~~~ backtrack{\U}stack.push(L{\U}CONT{\U}\m{\ruleNo});
\end{tabular}
\end{tt}
\end{quote}
This ensures that the \code{while}-loop above will be continued later.
Since 
then
the values of the variables
introduced in the rule will change,
this label must be on the stack below
every task using 
the generated tuple.
\item
Let \m{\ruleNo_1\until \ruleNo_k} be all rules
with an IDB body literal~\m{\litB_i}, \m{i\defEq 1\until k},
which matches the generated symbolic fact
\m{\pred(\bar\term_1\until\bar\term_n)}.
For \m{i\defEq 2\until k},
generate
\begin{quote}
\begin{tt}
\begin{tabular}{@{}l@{}}
~~~ ~~~ backtrack{\U}stack.push(%
	L{\U}START(\m{\pred(\bar\term_1\until\bar\term_n)},~\m{\ruleNo_i},~%
			\m{\litB_i}));
\end{tabular}
\end{tt}
\end{quote}
Finally,
generate
\begin{quote}
\begin{tt}
\begin{tabular}{@{}l@{}}
~~~ ~~~ goto
	l{\U}start(\m{\pred(\bar\term_1\until\bar\term_n)},~\m{\ruleNo_1},~%
	\m{\litB_1});
\end{tabular}
\end{tt}
\end{quote}
\item
Generate
\vspace*{2pt}
\begin{quote}
\begin{tt}
\begin{tabular}{@{}l@{}}
~~~ {\RB} // End of while-loop\\[-1pt]
~~~ break;
\end{tabular}
\end{tt}
\end{quote}
The \code{break;} is important if the \code{while}-loop ends
because no (further) matching fact is found in the relation~\m{\predC}.
Otherwise,
the loop is left with \code{goto}
when the first/next matching fact is found. 
\end{enumerate}

\subsection{Two EDB-Body Literals}

In the output of SLDMagic,
this case does not occur.
However,
it is easy to extend the above program code.
One uses two cursors,
one for each body literal,
and two nested \code{while}-loops.
For simplicity,
we implement all joins as ``nested loop join''
(or ``index join'' if the data structure for the relation
supports the corresponding binding pattern).
Later,
sort orders might be used,
so that also a ``merge join'' can be generated.

\subsection{One IDB-Body Literal}

\label{subsecOneIDBLit}

Consider the rule
\begin{quote}
\m{\pred(\term_1\until\term_n)\lif\predB(\termB_1\until\termB_m)},
\end{quote}
where \m{\predB} is an IDB-predicate.
Let \m{\ruleNo} be the number of this rule.
Due to partial evaluation done at compile time,
several specializations of the same rule might be generated.
There is one code piece
per symbolic fact \m{\predB(\bar\termB_1\until\bar\termB_m)}
which matches the body literal.
Let \m{\subst} be a most general unifier,
where variable-to-variable bindings are done such that
logic variables are replaced by~\CPP~variables
i.e.~\m{\termB_i} is replaced by~\m{\bar\termB_i},
if both are variables.
The generated symbolic fact is
\m{\pred(\doSubst{\subst}{\term_1}\until\doSubst{\subst}{\term_n})}.
Note that because of the range restriction requirement,
every variable among the~\m{\term_i}
also appears as an~\m{\termB_j},
and then it is unified with a constant or a \CPP~variable.
Thus,
no new \CPP\ variables are introduced in this case.
\begin{enumerate}
\item
Generate
\begin{quote}
\begin{tt}
\begin{tabular}{@{}l@{}}
case
L{\U}START(\m{\predB(\bar\termB_1\until\bar\termB_m)},~\m{\ruleNo},~%
	\m{\predB(\termB_1\until\termB_m)}):\\
l{\U}start(\m{\predB(\bar\termB_1\until\bar\termB_m)},~\m{\ruleNo},~%
	\m{\predB(\termB_1\until\termB_m)}):\\
\end{tabular}
\end{tt}
\end{quote}
If this rule application is activated via backtracking,
the \code{case} label is used.
If it is activated as the first usage of a generated fact,
a jump to the \code{goto} label is done
(as a slight optimization of pushing something on the backtrack stack
and immediately popping it again).
\item
\label{OneIDBBodyIf}%
Now the part of the unification
which can only be done at runtime must be generated.
Let \m{V_1\until V_k} be all \CPP~variables
which \m{\subst} replaces by constants
or a different variable
(i.e.~\m{\doSubst{\subst}{V_i}\ne V_i}).
If \m{k\gt 0}, generate:
\begin{quote}
\begin{tt}
\begin{tabular}{@{}l@{}}
~~~ if(\m{V_1} != \m{\doSubst{\subst}{V_1}}
	|| \m{\cdots} ||
	\m{V_k} != \m{\doSubst{\subst}{V_k}})\\
~~~ ~~~ break;
\end{tabular}
\end{tt}
\end{quote}
So we simply stop executing this code piece
if the current fact for~\m{\predB}
does not unify with the body literal.
Then another task will be taken from the backtrack stack
in the main loop.
\item
Next, if the predicate~\m{\pred} was selected for a duplicate check,
the code to enter the result tuple
\m{\pred(\doSubst{\subst}{\term_1}\until\doSubst{\subst}{\term_n})}
with the current values of the \CPP~variables
into a hash table is generated here.
If the tuple was already present
(so we just computed a duplicate),
one simply does ``\code{break;}''
to end the code piece
(as under \ref{OneIDBBodyIf} above).
\item
Let \m{\ruleNo_1\until \ruleNo_k} be all rules
with an IDB body literal~\m{\litB_i}, \m{i\defEq 1\until k},
which matches the generated symbolic fact
\m{\pred(\doSubst{\subst}{\term_1}\until\doSubst{\subst}{\term_n})}.
For \m{i\defEq 2\until k},
generate
\begin{quote}
\begin{tt}
\begin{tabular}{@{}l@{}}
~~~ backtrack{\U}stack.push(%
	L{\U}START(\m{\pred(\doSubst{\subst}{\term_1}\until
			\doSubst{\subst}{\term_n})},~\m{\ruleNo_i},~%
				\m{\litB_i}));
\end{tabular}
\end{tt}
\end{quote}
Finally,
generate
\begin{quote}
\begin{tt}
\begin{tabular}{@{}l@{}}
~~~ goto l{\U}start(\m{\pred(\doSubst{\subst}{\term_1}\until
			\doSubst{\subst}{\term_n})},~%
			\m{\ruleNo_1},~\m{\litB_1});
\end{tabular}
\end{tt}
\end{quote}
\end{enumerate}

\subsection{One IDB- and one EDB-Body Literal}

Consider the rule
\begin{quote}
\m{\pred(\term_1\until\term_n)\lif\predB(\termB_1\until\termB_m)\land
	\predC(\termC_1\until\termC_l)},
\end{quote}
where \m{\predB} is an IDB-predicate
and \m{\predC} is an EDB-predicate.
Let \m{\ruleNo} be the number of this rule.
Again,
there is one code piece
per symbolic fact \m{\predB(\bar\termB_1\until\bar\termB_m)}
which matches the IDB body literal.
Let \m{\subst} be a most general unifier
and \m{\pred(\bar\term_1\until\bar\term_n)}
be the generated symbolic fact
as defined in Section~\ref{subsecRuleApplicationGraph}.
As in Section~\ref{subsecOneEDBLit},
select a binding pattern~\m{\binding}
for accessing the EDB-relation~\m{\predC}.
A value for~\m{\termC_i} is known
(i.e.~\m{\binding_i} can be ``bound'')
if \m{\termC_i} is a constant
or a variable which also appears in~\m{\predB(\termB_1\until\termB_m)}
(when execution reaches this code piece,
a concrete fact is given for the IDB body literal).
In the declaration section, generate
\begin{quote}
\begin{tt}
\begin{tabular}{@{}l@{}}
cursor{\U}\m{\predC}{\U}\m{\binding} c\m{\ruleNo};\\
\end{tabular}
\end{tt}
\end{quote}
All following code is generated in the \code{switch}:
\begin{enumerate}
\item
Generate
\begin{quote}
\begin{tt}
\begin{tabular}{@{}l@{}}
case
L{\U}START(\m{\predB(\bar\termB_1\until\bar\termB_m)},~\m{\ruleNo},~%
	\m{\predB(\termB_1\until\termB_m)}):\\
l{\U}start(\m{\predB(\bar\termB_1\until\bar\termB_m)},~\m{\ruleNo},~%
	\m{\predB(\termB_1\until\termB_m)}):\\
\end{tabular}
\end{tt}
\end{quote}
\item
Now the part of the unification
of the given fact with the IDB body literal,
which can only be done at runtime,
must be generated.
Let \m{V_1\until V_k} be all \CPP~variables
with \m{\doSubst{\subst}{V_i}\ne V_i}.
If \m{k\gt 0}, generate:
\begin{quote}
\begin{tt}
\begin{tabular}{@{}l@{}}
~~~ if(\m{V_1} != \m{\doSubst{\subst}{V_1}}
	|| \m{\cdots} ||
	\m{V_k} != \m{\doSubst{\subst}{V_k}})\\
~~~ ~~~ break;
\end{tabular}
\end{tt}
\end{quote}
This ends the execution of this code piece
if the rule is not applicable.
\item
\label{IDBEDBRecursion}
Now we must use a cursor
to access the tuples for
the EDB body literal~\m{\predC(\termC_1\until\termC_l)}.
In case this rule is recursive,
it might be possible
that the state of the cursor and
the values of the variables~\code{v\m{\ruleNo}{\U}\m{\varA}}
set in this rule
are still needed
by backtrack points on the stack
(unless we know 
that there are 
no such backtrack points,
e.g.~because the recursive rule application
is the last use of the fact).
Therefore,
we generate
\begin{quote}
\begin{tt}
\begin{tabular}{@{}l@{}}
~~~ c\m{\ruleNo}.push();\\
\end{tabular}
\end{tt}
\end{quote}
And for each variable~\code{v{\m\ruleNo}{\U}\m{\varA}}
we generate
\begin{quote}
\begin{tt}
\begin{tabular}{@{}l@{}}
~~~ value{\U}stack.push(v\m{\ruleNo}{\U}\m{\varA});
\end{tabular}
\end{tt}
\end{quote}
(There are probably several value stacks for different data types.)\\
Finally we generate a backtrack point:
\begin{quote}
\begin{tt}
\begin{tabular}{@{}l@{}}
~~~ backtrack{\U}stack.push(L{\U}RESTORE{\U}\m{\ruleNo});
\end{tabular}
\end{tt}
\end{quote}
The code for this \code{case} in the \code{switch}
simply restores the variable values and the cursor state
by popping them (in the inverse order).
In this way,
all earlier backtrack points
(below the one just generated)
find the old cursor state and variable values.
\item
\label{IDBEDBLitKnownArg}
Now we open the cursor over the EDB-relation~\m{\predC}
with the selected binding pattern~\m{\binding}.
Let \m{i_1\until i_k} be the bound argument positions in~\m{\binding}.
Generate:
\begin{quote}
\begin{tt}
\begin{tabular}{@{}l@{}}
~~~ c\m{\ruleNo}.open(\m{\bar\termC_{i_1}\until\bar\termC_{i_k}});
\end{tabular}
\end{tt}
\end{quote}
where \m{\bar\termC_{i_j}} is
\begin{itemize}
\item
\m{\termC_{i_j}} if this is a constant,
\item
\m{\doSubst{\subst}{\termC_{i_j}}} if
\m{\termC_{i_j}} is a variable
which appears in the IDB body literal~\m{\predB(\ldots)}.
\end{itemize}
\item
Generate:
\begin{quote}
\begin{tt}
\begin{tabular}{@{}l@{}}
case L{\U}CONT{\U}\m{\ruleNo}:
\end{tabular}
\end{tt}
\end{quote}
When we are finished with using the computed fact,
backtracking returns here to continue the following loop.
\item
\label{IDBEDBLitWhile}
Generate
\begin{quote}
\begin{tt}
\begin{tabular}{@{}l@{}}
~~~ while(c\m{\ruleNo}.fetch()) {\LB}
\end{tabular}
\end{tt}
\end{quote}
\item
Now we do that part of the section for~\m{\predC(\termC_1\until\termC_l)}
that was not supported by the data structure for~\m{\predC}
with binding pattern~\m{\binding} (e.g., hash table).
Let \m{i_1\until i_k} be the free argument positions in~\m{\binding}
such that \m{\termC_{i_j}} is a constant
or a variable which appears the the IDB body literal~\m{\predB(\ldots)}.
Let \m{\bar\termC_{i_j}} be defined as in~\ref{IDBEDBLitKnownArg} above.
If \m{k\ge1},
generate
\begin{quote}
\begin{tt}
\begin{tabular}{@{}l@{}}
~~~ ~~~ if(c\m{\ruleNo}.col{\U}\m{i_1}() != \m{\bar\termC_{i_1}}
	|| \m{\cdots} ||
	c\m{\ruleNo}.col{\U}\m{i_k}() != \m{\bar\termC_{i_k}})\\
~~~ ~~~ ~~~ continue;
\end{tabular}
\end{tt}
\end{quote}
I.e.~if the current tuple in the EDB relation
does not have the required values,
we continue with the next iteration of the \code{while}-loop
under~\ref{IDBEDBLitWhile}.
\item
For every variable~\m{\varB},
which appears more than once among the~\m{\termC_1\until\termC_l},
but not in the IDB~literal~\m{\predB(\ldots)}:
Let \m{\termB_{i_1}\until\termB_{i_k}}
be all equal to~\m{\varB}
(note that \m{k\ge 2}).
Generate:
\begin{quote}
\begin{tt}
\begin{tabular}{@{}l@{}}
~~~ ~~~ if(c\m{\ruleNo}.col{\U}\m{i_1}() \kern-2pt != \kern-2pt
			c\m{\ruleNo}.col{\U}\m{i_2}
	\kern-2pt || \kern-2pt \m{\cdots} \kern-2pt || \kern-2pt
	c\m{\ruleNo}.col{\U}\m{i_{k-1}}() \kern-2pt != \kern-2pt
			c\m{\ruleNo}.col{\U}\m{i_k}())\kern-17pt\\
~~~ ~~~ ~~~ continue;
\end{tabular}
\end{tt}
\end{quote}
\item
For every variable~\m{\varA_i} in the head,
which does not appear in the IDB body literal~\m{\predB(\ldots)},
let \m{\termC_j} be any occurrence of this variable
among the \m{\termC_1\until\termC_l}.
Because of the range restriction (allowedness) condition on the rules,
it must occur there.
Generate for each~\m{\varA_i}:
\begin{quote}
\begin{tt}
\begin{tabular}{@{}l@{}}
~~~ ~~~ v\m{\ruleNo}{\U}\m{\varA_i} = c\m{\ruleNo}.col{\U}\m{j}();
\end{tabular}
\end{tt}
\end{quote}
\item
In case predicate~\m{\pred} was selected for a duplicate check,
we again enter the result tuple
\m{\pred(\bar\term_1\until\bar\term_n)}
with the current values of the \CPP~variables
into a hash table. 
If the tuple was already present,
one simply does ``\code{continue;}''
to compute the next tuple.
\item
Generate:
\begin{quote}
\begin{tt}
\begin{tabular}{@{}l@{}}
~~~ ~~~ backtrack{\U}stack.push(L{\U}CONT{\U}\m{\ruleNo});
\end{tabular}
\end{tt}
\end{quote}
\item
Let \m{\ruleNo_1\until \ruleNo_k} be all rules
with an IDB body literal~\m{\litB_i}, \m{i\defEq 1\until k}
that matches the generated symbolic fact
\m{\pred(\bar\term_1\until\bar\term_n)}.
For \m{i\defEq 2\until k},
generate
\begin{quote}
\begin{tt}
\begin{tabular}{@{}l@{}}
~~~ ~~~ backtrack{\U}stack.push(%
	L{\U}START(\m{\pred(\bar\term_1\until\bar\term_n)},~\m{\ruleNo_i},~%
	\m{\litB_i}));
\end{tabular}
\end{tt}
\end{quote}
Then
generate
\begin{quote}
\begin{tt}
\begin{tabular}{@{}l@{}}
~~~ ~~~ goto
	l{\U}start(\m{\pred(\bar\term_1\until\bar\term_n)},~\m{\ruleNo_1},~%
		\m{\litB_1});
\end{tabular}
\end{tt}
\end{quote}
\item
Finally,
we must close the open \code{while}-loop (\ref{IDBEDBLitWhile}.~above)
and finish the code piece
in case the loop does not find any (further) matching tuple.
Generate:
\begin{quote}
\begin{tt}
\begin{tabular}{@{}l@{}}
~~~ {\RB}\\
~~~ break;
\end{tabular}
\end{tt}
\end{quote}
\end{enumerate}
In addition,
there is a code piece for \code{case~L{\U}RESTORE{\U}\m{\ruleNo}}
as explained under item~\ref{IDBEDBRecursion} above.
It is executed when this rule application is finished.
It pops everything pushed there
and then does \code{break;}
to continue with the next task from the backtrack stack.

\subsection{Two IDB-Body Literals}

This is a complicated case
and needs intermediate storage of tuples generated
for the body literals.
Fortunately,
this occurs rarely 
in the output of the SLDMagic method
(only when translating recursions that are not tail recursions).

A general solution,
which does not need information about the order of generated tuples,
is to manage one set of tuples for each body literal.
In the code piece
for the case that a new tuple has been derived for the left body literal,
this tuple is joined with all tuples
in the current set for the right body literal.
In the same way,
when a new tuple is generated for the right body literal,
it is joined with all existing tuples for the left body literal.

This means that we now need cursors also for the intermediate storage
of generated IDB facts,
and these cursors must keep information about the last fact
when they were created
(since new facts can be appended to the list
while the cursor is active---these 
facts must not be returned by the cursor).

Recursion can be handled in the same way as before:
When a new fact is generated for a body literal,
we save the state of the cursor and all variables of that code piece,
continue with derivations using the new fact,
and later return to the old fact.
However,
since we anyway have intermediate storage now,
it is also possible to create a queue of facts for each body literal,
which must still be used in derivations.

If it is possible to generate 
all facts for the left body literal
before the first fact for the right body literal,
one obviously needs intermediate storage only for the left body literal.
In this case it can later be treated like an EDB literal.

If one should be able
generate facts for both literals in the sort order
of the common variables (the join attributes),
one would need intermediate storage only for a single tuple
for each body literal (we would basically do a merge join).

\subsection{Generated Answer-Facts}

For rules about the predicate \m{\answerPred},
one can print the generated tuple
or insert it into a result relation
whenever the above code would jump to a body literal
which uses the generated fact
(there are no body literals with predicate~\m{\answerPred}).
One can also offer
the cursor interface of Section~\ref{secDatabaseInterface}.

\section{Conclusion}

We have presented a detailed description of the push method,
an efficient bottom-up evaluation algorithm for pure Datalog programs.
It is implemented as a translation from Datalog to \CPP.

In the push method, 
derived facts are immediately used to derive new facts without generally 
materializing immediate results.
A specific feature introduced in this paper
is the representation of derived tuples
which significantly reduces the amount of copying.
This is the result of the partial evaluation
to do as much as possible work already at ``compilation time''
(at the price of producing several specializations of the same rule,
i.e.~the generated code might grow while runtime is saved).
The rule application graph defined here is useful for planning
the evaluation.

First performance tests show some improvement over the previous version
of the push method from~\cite{Bra10CPP}.
We plan to develop a more complete implementation
and to investigate further optimizations.
The current state of the project is reported at
\begin{quote}
\url{http://www.informatik.uni-halle.de/~brass/push/}.
\end{quote}

\bibliographystyle{eptcs}

\bibliography{ddb}

\begin{thebibliography}{10}
\providecommand{\bibitemdeclare}[2]{}
\providecommand{\surnamestart}{}
\providecommand{\surnameend}{}
\providecommand{\urlprefix}{Available at }
\providecommand{\url}[1]{\texttt{#1}}
\providecommand{\href}[2]{\texttt{#2}}
\providecommand{\urlalt}[2]{\href{#1}{#2}}
\providecommand{\doi}[1]{doi:\urlalt{http://dx.doi.org/#1}{#1}}
\providecommand{\bibinfo}[2]{#2}

\bibitemdeclare{inproceedings}{Bra00}
\bibitem{Bra00}
\bibinfo{author}{Stefan \surnamestart Brass\surnameend} (\bibinfo{year}{2000}):
  \emph{\bibinfo{title}{{SLDMagic} --- The Real Magic (with Applications to Web
  Queries).}}
\newblock In \bibinfo{editor}{W.~\surnamestart Lloyd\surnameend} et~al.,
  editors: {\sl \bibinfo{booktitle}{First International Conference on
  Computational Logic (CL'2000/DOOD'2000)}}, {\sl \bibinfo{series}{LNCS}}
  \bibinfo{volume}{1861}, \bibinfo{publisher}{Springer}, pp.
  \bibinfo{pages}{1063--1077}, \doi{10.1007/3-540-44957-4\_71}.
\newblock
  \urlprefix\url{http://users.informatik.uni-halle.de/~brass/sldmagic/}.

\bibitemdeclare{inproceedings}{Bra10CPP}
\bibitem{Bra10CPP}
\bibinfo{author}{Stefan \surnamestart Brass\surnameend} (\bibinfo{year}{2010}):
  \emph{\bibinfo{title}{Implementation Alternatives for Bottom-Up Evaluation}}.
\newblock In \bibinfo{editor}{Manuel \surnamestart Hermenegildo\surnameend} \&
  \bibinfo{editor}{Torsten \surnamestart Schaub\surnameend}, editors: {\sl
  \bibinfo{booktitle}{Technical Communications of the 26th International
  Conference on Logic Programming (ICLP'10)}}, {\sl \bibinfo{series}{Leibniz
  International Proceedings in Informatics (LIPIcs)}}~\bibinfo{volume}{7},
  \bibinfo{publisher}{Schloss Dagstuhl}, pp. \bibinfo{pages}{44--53},
  \doi{10.4230/LIPIcs.ICLP.2010.44}.
\newblock \urlprefix\url{http://users.informatik.uni-halle.de/~brass/botup/}.

\bibitemdeclare{inproceedings}{BS13}
\bibitem{BS13}
\bibinfo{author}{Stefan \surnamestart Brass\surnameend} \&
  \bibinfo{author}{Heike \surnamestart Stephan\surnameend}
  (\bibinfo{year}{2013}): \emph{\bibinfo{title}{A Variant of Earley Deduction
  with Partial Evaluation}}.
\newblock In \bibinfo{editor}{Wolfgang \surnamestart Faber\surnameend} \&
  \bibinfo{editor}{Domenico \surnamestart Lembo\surnameend}, editors: {\sl
  \bibinfo{booktitle}{Web Reasoning and Rule Systems - 7th International
  Conference, RR 2013}}, {\sl \bibinfo{series}{LNCS}} \bibinfo{volume}{7994},
  \bibinfo{publisher}{Springer-Verlag}, pp. \bibinfo{pages}{35--49},
  \doi{10.1007/978-3-642-39666-3\_4}.
\newblock \urlprefix\url{http://dbs.informatik.uni-halle.de/Earley/}.

\bibitemdeclare{article}{DMP94}
\bibitem{DMP94}
\bibinfo{author}{Marcia~A. \surnamestart Derr\surnameend},
  \bibinfo{author}{Shinichi \surnamestart Morishita\surnameend} \&
  \bibinfo{author}{Geoffrey \surnamestart Phipps\surnameend}
  (\bibinfo{year}{1994}): \emph{\bibinfo{title}{The {G}lue-{N}ail Deductive
  Database System: Design, Implementation and Evaluation}}.
\newblock {\sl \bibinfo{journal}{The VLDB Journal}} \bibinfo{volume}{3}, pp.
  \bibinfo{pages}{123--160}, \doi{10.1007/BF01228879}.
\newblock \urlprefix\url{http://www.vldb.org/journal/VLDBJ3/P123.pdf}.

\bibitemdeclare{inproceedings}{GAK12}
\bibitem{GAK12}
\bibinfo{author}{Todd~J. \surnamestart Green\surnameend},
  \bibinfo{author}{Molham \surnamestart Aref\surnameend} \&
  \bibinfo{author}{Grigoris \surnamestart Karvounarakis\surnameend}
  (\bibinfo{year}{2012}): \emph{\bibinfo{title}{{LogicBlox}, Platform and
  Language: A Tutorial}}.
\newblock In \bibinfo{editor}{Pablo \surnamestart Barcel{\'o}\surnameend} \&
  \bibinfo{editor}{Reinhard \surnamestart Pichler\surnameend}, editors: {\sl
  \bibinfo{booktitle}{Datalog in Academica and Industry, 2nd Int.~Workshop,
  Datalog~2.0}}, {\sl \bibinfo{series}{LNCS}} \bibinfo{volume}{7494},
  \bibinfo{publisher}{Springer-Verlag}, pp. \bibinfo{pages}{1--8},
  \doi{10.1007/978-3-642-32925-8\_1}.
\newblock \urlprefix\url{https://developer.logicblox.com/2012/08/}.

\bibitemdeclare{inproceedings}{LFWK09}
\bibitem{LFWK09}
\bibinfo{author}{Senlin \surnamestart Liang\surnameend}, \bibinfo{author}{Paul
  \surnamestart Fodor\surnameend}, \bibinfo{author}{Hui \surnamestart
  Wan\surnameend} \& \bibinfo{author}{Michael \surnamestart Kifer\surnameend}
  (\bibinfo{year}{2009}): \emph{\bibinfo{title}{{OpenRuleBench}: {An} Analysis
  of the Performance of Rule Engines}}.
\newblock In: {\sl \bibinfo{booktitle}{Proceedings of the 18th International
  Conference on World Wide Web (WWW'09)}}, \bibinfo{publisher}{ACM}, pp.
  \bibinfo{pages}{601--610}, \doi{10.1145/1526709.1526790}.
\newblock \urlprefix\url{http://rulebench.projects.semwebcentral.org/}.

\bibitemdeclare{article}{Liu99Rol}
\bibitem{Liu99Rol}
\bibinfo{author}{Mengchi \surnamestart Liu\surnameend} (\bibinfo{year}{2000}):
  \emph{\bibinfo{title}{Design and Implementation of the {ROL} System}}.
\newblock {\sl \bibinfo{journal}{Journal of Intelligent Information Systems}}
  \bibinfo{volume}{14}, pp. \bibinfo{pages}{1--21},
  \doi{10.1023/A:1008774006482}.
\newblock
  \urlprefix\url{http://www.scs.carleton.ca/~mengchi/papers/rol-JIIS00.ps}.

\bibitemdeclare{inproceedings}{Ram93}
\bibitem{Ram93}
\bibinfo{author}{Kotagiri \surnamestart Ramamohanarao\surnameend}
  (\bibinfo{year}{1993}): \emph{\bibinfo{title}{An Implementation Overview of
  the {A}diti Deductive Database System}}.
\newblock In \bibinfo{editor}{Stefano \surnamestart Ceri\surnameend},
  \bibinfo{editor}{Katsumi \surnamestart Tanaka\surnameend} \&
  \bibinfo{editor}{Shalom \surnamestart Tsur\surnameend}, editors: {\sl
  \bibinfo{booktitle}{Deductive and Object-Oriented Databases, Third
  Int.~Conf.,~(DOOD'93)}}, {\sl \bibinfo{series}{LNCS}} \bibinfo{volume}{760},
  \bibinfo{publisher}{Springer}, pp. \bibinfo{pages}{184--203},
  \doi{10.1007/3-540-57530-8\_12}.

\bibitemdeclare{inproceedings}{SSW94}
\bibitem{SSW94}
\bibinfo{author}{Konstantinos \surnamestart Sagonas\surnameend},
  \bibinfo{author}{Terrance \surnamestart Swift\surnameend} \&
  \bibinfo{author}{David~S. \surnamestart Warren\surnameend}
  (\bibinfo{year}{1994}): \emph{\bibinfo{title}{{XSB} as an Efficient Deductive
  Database Engine}}.
\newblock In \bibinfo{editor}{Richard~T. \surnamestart Snodgrass\surnameend} \&
  \bibinfo{editor}{Marianne \surnamestart Winslett\surnameend}, editors: {\sl
  \bibinfo{booktitle}{Proc.~of the 1994 ACM SIGMOD Int.~Conf.~on Management of
  Data (SIGMOD'94)}}, pp. \bibinfo{pages}{442--453},
  \doi{10.1145/191843.191927}.
\newblock \urlprefix\url{http://user.it.uu.se/~kostis/Papers/xsbddb.html}.

\bibitemdeclare{phdthesis}{Sch93}
\bibitem{Sch93}
\bibinfo{author}{Heribert \surnamestart Sch{\"u}tz\surnameend}
  (\bibinfo{year}{1993}): \emph{\bibinfo{title}{{T}upelweise
  {B}ottom-up-{A}uswertung von {L}ogikprogrammen (Tuple-wise bottom-up
  evaluation of logic programs)}}.
\newblock Ph.D. thesis, \bibinfo{school}{TU M{\"u}nchen}.

\bibitemdeclare{techreport}{SFH96}
\bibitem{SFH96}
\bibinfo{author}{Praveen \surnamestart Seshadri\surnameend},
  \bibinfo{author}{Shaun \surnamestart Flisakowski\surnameend} \&
  \bibinfo{author}{Seymour \surnamestart Hersh\surnameend}
  (\bibinfo{year}{1996}): \emph{\bibinfo{title}{CORAL: The Inside Story.
  {Shocking} Hacks Revealed}}.
\newblock \bibinfo{type}{Technical Report}, \bibinfo{institution}{Department of
  Computer Sciences, The University of Wisconsin-Madison}.
\newblock \urlprefix\url{http://ftp.cs.wisc.edu/coral/doc/Inside.ps}.

\bibitemdeclare{inproceedings}{SU99}
\bibitem{SU99}
\bibinfo{author}{Donald~A. \surnamestart Smith\surnameend} \&
  \bibinfo{author}{Mark \surnamestart Utting\surnameend}
  (\bibinfo{year}{1999}): \emph{\bibinfo{title}{Pseudo-Naive Evaluation}}.
\newblock In: {\sl \bibinfo{booktitle}{Australasian Database Conference}}, pp.
  \bibinfo{pages}{211--223}.
\newblock
  \urlprefix\url{http://citeseerx.ist.psu.edu/viewdoc/summary?doi=10.1.1.177.5047}.
\newblock \bibinfo{note}{\\{\tt http://www.cs.waikato.ac.nz/research/jstar/
  1999-ADC-pseudo-naive-eval.pdf}}.

\bibitemdeclare{inproceedings}{YK00}
\bibitem{YK00}
\bibinfo{author}{Guizhen \surnamestart Yang\surnameend} \&
  \bibinfo{author}{Michael \surnamestart Kifer\surnameend}
  (\bibinfo{year}{2000}): \emph{\bibinfo{title}{{FLORA}: Implementing an
  Efficient {DOOD} System Using a Tabling Logic Engine}}.
\newblock In \bibinfo{editor}{W.~\surnamestart Lloyd\surnameend} et~al.,
  editors: {\sl \bibinfo{booktitle}{First International Conference on
  Computational Logic (CL'2000/DOOD'2000)}}, {\sl \bibinfo{series}{LNCS}}
  \bibinfo{volume}{1861}, \bibinfo{publisher}{Springer}, pp.
  \bibinfo{pages}{1078--1093}, \doi{10.1007/3-540-44957-4\_72}.
\newblock
  \urlprefix\url{http://citeseerx.ist.psu.edu/viewdoc/summary?doi=10.1.1.110.4304}.

\end{thebibliography}

\end{document}